\documentclass [a4paper,fleqn, 12pt]{article}
\usepackage{graphicx}

\usepackage {amsmath} \usepackage{amssymb} \usepackage{cite}

\newcommand{\cn}

\begin{document}

\title{The fifth-order partial differential equation for the description of the $\alpha+\beta$ Fermi--Pasta--Ulam model}

\author{Nikolay A. Kudryashov, \and Alexandr K. Volkov}

\date{Department of Applied Mathematics, National Research Nuclear University MEPHI, 31 Kashirskoe Shosse, 115409 Moscow, Russian Federation}

\maketitle

\begin{abstract}
We study a new nonlinear partial differential equation of the fifth order for the description of perturbations in the Fermi-Pasta-Ulam mass chain. This fifth-order equation is an expansion of the Gardner equation for the description of the Fermi--Pasta--Ulam model. We use the potential of interaction between neighbouring masses with both quadratic and cubic terms. The equation is derived using the continuous limit. Unlike the previous works, we take into account  higher order terms in the Taylor series expansions. We  investigate the equation using the Painlev\'e approach. We show that the equation does not pass the Painlev\'e test and can not be integrated by the inverse scattering transform. We use the logistic function method and the Laurent expansion method to find travelling wave solutions of the fifth--order equation. We use the pseudospectral method for the numerical simulation of wave processes, described by the equation.
\end{abstract}


\section{Introduction}
The Fermi--Pasta--Ulam (FPU) model was first studied in work \cite{Fermi}. It describes perturbations in the chain of nonlinear coupling among masses. It is shown \cite{Fermi} that energy remains in a very few modes and the long-time dynamic of the system is recurrent.
This fact was called the FPU paradox. A set of approaches was used to explain it (see for example \cite{Ford1961, izrailev1970, wu2007, livi1985, Ruffo, porubov2014non}). The main ideas of these approaches and their results can be found in work \cite{Berman}.

Usually $\alpha$ or $\beta$ FPU model is considered in literature, but in paper \cite{Benettin2013} the $\alpha+\beta$ FPU model is investigated. In this paper, we also use the $\alpha+\beta$ model but we use the continuous limit approximation for the investigation. 

This approach was used for the first time by N. Zabusky and M. Kruskal in work \cite{Kruskal}, where the authors derived the Korteweg--de Vries (KdV) equation for the description of the $\alpha$ FPU model. This equation is integrable by the inverse scattering transform and has been widely studied by now (\cite{Kudr_book, biswas}). Zabusky and Kruskal explained the FPU paradox (recurrence of initial conditions) using the numerical simulation of the KdV equation. Another result of work \cite{Kruskal} was an introduction of soliton. In article \cite{KudrPLA15} the author took into account high order (in comparison with \cite{Kruskal}) terms in the Taylor series expansions for the continuous limit approximation of the $\alpha$ FPU model. The generalized KdV equation was obtained for a more accurate description of wave processes in the FPU model. Some exact solutions of the derived equation were found . It is shown \cite{KudrPLA15} that if one takes into account more terms in the Taylor series and acts in analogy to paper \cite{Kruskal} then he can not derive an integrable equation for the description of wave processes in the FPU mass chain. Recurrence of initial statement is also unrepresentative for wave processes, described by the fifth-order equation with arbitrary initial conditions. In paper \cite{Volkov} wave processes for the generalized KdV equation are simulated numerically. 

This brings up the question, whether the situation is similar for the $\beta$ and the $\alpha + \beta$ FPU models? It is known \cite{Kudr_book} that one can obtain the Gardner equation for the description of the $\alpha + \beta$ FPU model using the continuous limit. The Gardner equation is integrable by the inverse scatterring transform and can be reduced to the modified KdV equation \cite{Kudryashov2011note}. If we take more terms in the Taylor series expansion, we can derive the fifth-order partial differential equation for the description of the $\alpha+\beta$ FPU model. To the best of our knowledge, this equation has not been obtained and investigated before. This is the main aim of the current manuscript.

The rest of the work is organized as follows. In section 2 we derive the fifth-order partial differential equation for the description of the $\alpha + \beta$ FPU model. In section 3 we use the Painlev\'e test to analyze the derived equation. The logistic function method is used in section 4 to find exact solutions of the derived equation. In section 5 we obtain an elliptic solution for this equation. In section 6 we present the results for the numerical simulation of wave processes, described by the derived equation. In section 7 we briefly discuss our results. 


\section{The fifth-order partial differential equation for the description of the $\alpha+\beta$ Fermi--Pasta--Ulam model}

Let us consider the $\alpha+\beta$ Fermi--Pasta--Ulam model. It appears as the following system of equations:
\begin{equation}
\begin{gathered}
\label{1d_dicrete}
m\frac{d^2y_i}{d\tau^2} = \gamma(y_{i+1} - y_i) + \alpha(y_{i+1} - y_i)^2 - \beta(y_{i+1} - y_i)^3 - \\
- \gamma(y_i - y_{i-1}) - \alpha( y_i - y_{i-1})^2 + \beta(y_i - y_{i-1})^3 = \\
= (y_{i+1} -2y_i + y_{i-1})
(\gamma +\alpha (y_{i+1} - y_{i-1}) - \\
- \beta(y_{i+1}^2 + y_{i}^2 + y_{i-1}^2 -y_iy_{i+1} - y_iy_{i-1} - y_{i+1}y_{i-1})),
\end{gathered}
\end{equation}
where $y_i$ is a displacement of the particle number $i$ from its equilibrium position, $i$ runs through values from $1$ to $N$, where $N$ is the number of particles, $m$ is the mass of a single particle, $\alpha,\; \beta,\; \gamma$ are some positive constants that characterize the potential of interaction and $\tau$ is the time. Let us find the continuous limit approximation of the model. In order to do this we use the following Taylor series expansions: 
\begin{equation}
\begin{gathered}
\label{1d_taylor}
y_{i\pm1} = y \pm hy_{\xi} + \frac{h^2}{2}y_{\xi\xi} \pm \frac{h^3}{6}y_{\xi\xi\xi} + \frac{h^4}{24}y_{\xi\xi\xi\xi} \pm \\
\pm \frac{h^5}{120}y_{\xi\xi\xi\xi\xi} + \frac{h^6}{720}y_{\xi\xi\xi\xi\xi\xi}+\ldots,
\end{gathered}
\end{equation}
where $y=y(\xi, \tau)$ is a displacement of the infinitely small part of the chord with coordinate $\xi$ in time moment $\tau$, $h$ is a small parameter. Substituting expansion \eqref{1d_taylor} into the system of equations \eqref{1d_dicrete} we derive a partial differential equation for the description of the system dynamics in the continuous limit approximation. Let us take into account only terms up to $h^6$.

\begin{equation}
\begin{gathered}
\label{1d_contin_h}
m\frac{d^2y}{d\tau^2} = h^2\gamma y_{\xi\xi} +2\alpha h^3 y_{\xi}y_{\xi\xi} - 3\beta h^4y^2_{\xi}y_{\xi\xi} + \frac{\gamma h^4}{12}y_{\xi\xi\xi\xi} + 
+ \frac{\alpha h^5}{3}y_{\xi\xi}y_{\xi\xi\xi} +\\
+ \frac{\alpha h^5}{6}y_{\xi}y_{\xi\xi\xi\xi} - \beta h^6 y_{\xi}y_{\xi\xi}y_{\xi\xi\xi} 
- \frac{\beta h^6}{4}y^3_{\xi\xi} - \frac{\beta h^6}{4}y^2_{\xi}y_{\xi\xi\xi\xi} + \frac{\gamma h^6}{360}y_{\xi\xi\xi\xi\xi\xi}.
\end{gathered}
\end{equation}
Using a new parameter:
\begin{equation}
c^2 = \frac{\gamma h^2}{m},
\end{equation}
and scaling transformations:
\begin{equation}
\begin{gathered}
\label{1d_subst}
y=\frac{\gamma\sqrt{\gamma}}{2c\alpha \sqrt{m}}y',\qquad
T = \frac{c}{2}\tau,
\end{gathered}
\end{equation}
we reduce equation \eqref{1d_contin_h} to the form:
\begin{equation}
\begin{gathered}
\label{1d_contin_c}
\frac 14 \frac{d^2y}{dT^2} = y_{\xi\xi} + y_{\xi}y_{\xi\xi} - \mu y_{\xi}^2y_{\xi\xi} + 
\delta y_{\xi\xi\xi\xi} + 2\delta y_{\xi\xi}y_{\xi\xi\xi} + \delta y_{\xi}y_{\xi\xi\xi\xi} -\\
- 4\mu\delta y_{\xi}y_{\xi\xi}y_{\xi\xi\xi} - \mu \delta y_{\xi\xi}^3 - \mu \delta y_{\xi}^2y_{\xi\xi\xi\xi} + 
\frac{2}{5}\delta^2 y_{\xi\xi\xi\xi\xi\xi},
\end{gathered}
\end{equation}
where $\mu=\frac{3\beta\gamma}{4\alpha^2},\; \delta = \frac{mc^2}{12\gamma}$ and primes are omitted. 
According to \cite{Kudr_book}, the solution of this equation can be found as two oppositely directed running waves. Following \cite{Kruskal}, we only study the wave propagating to the right. For this purpose we search for the solution of equation \eqref{1d_contin_c} in the following form:
\begin{equation}
y(\xi, T) = f(x, t) + \varepsilon y_1(\xi, T),\quad
x = \xi - 2T,\quad t=\varepsilon T.
\end{equation}
Here $f(x, t) $ corresponds to the wave profile on long distance. We omit members that have high order in epsilon. Then from equation \eqref{1d_contin_c}, we derive an equation for $f(x, t)$:
\begin{equation}
\begin{gathered}
\label{1d_contin_f}
f_{xt} + f_{x}f_{xx} - \mu f_{x}^2f_{xx} +
\delta f_{xxxx} + 2\delta f_{xx}f_{xxx} + \delta f_{x}f_{xxxx} -\\
- 4\mu \delta f_{x}f_{xx}f_{xxx} - \mu \delta f_{xx}^3 - 
\mu \delta f_{x}^2f_{xxxx} + \frac{2}{5}\delta^2f_{xxxxxx} =0.
\end{gathered}
\end{equation}
Let us use a new variable:
\begin{equation}
u(x,t) = f_x(x,t),
\end{equation}
then equation \eqref{1d_contin_f} takes the form:
\begin{equation}
\begin{gathered}
\label{1d_final}
u_t + uu_x - \mu u^2u_x + \delta^2 u_{xxx} + 2\delta^2u_xu_{xx} + \delta^2uu_{xxx} -\\
- 4\delta^2\mu uu_xu_{xx} - \delta^2\mu u_x^3 - \delta^2\mu u^2u_{xxx} + \frac 25\delta^4 u_{xxxxx} = 0,
\end{gathered}
\end{equation}
Equation \eqref{1d_final} is the logical extension of the equation  from the article \cite{KudrPLA15}, because in case of quadratic potential of interaction ($\mu=0$), one can see that equation \eqref{1d_final} takes the form of equation, studied in \cite{KudrPLA15}. Parameter $\delta$ has also been introduced in the same form as in earlier studies of N. Zabusky \cite{Kruskal} and N. Kudryashov \cite{KudrPLA15}. Another type of the extended fifth-ordered KdV equations is considered in \cite{osborne}. Cases of the integrability for a lot of fifth-order PDEs can be found in \cite{kodama}. However, equation \eqref{1d_final} is new and worth being investigated.

If we use only terms up to the $h^4$ in \eqref{1d_contin_h}, we obtain the Gardner equation for the description of wave processes in the $\alpha+\beta$ FPU model:
\begin{equation}
\label{Gardner_eq}
u_t + uu_x - \mu u^2u_x + \delta^2 u_{xxx} = 0.
\end{equation}
As it was mentioned above,  equation \eqref{Gardner_eq} can be reduced to the  modified KdV equation. It is integrable and has solutions in the form of breather waves and solitons.


\section{Painlev\'e test for equation \eqref{1d_final}}
Let us use the Painlev\'e test for equation \eqref{1d_final} using the traveling wave variables. Below we will show that it is not necessary to analyse equation \eqref{1d_final} in general case. The traveling wave variables for equation \eqref{1d_final} have the form:
\begin{equation}
\label{1d_trav_wave_var}
u(x,t) = v(z),\qquad z=x-C_0t.
\end{equation}
Using \eqref{1d_trav_wave_var} we get from \eqref{1d_final}:
\begin{equation}
\begin{gathered}
\label{1d_trav_wave}
-C_0v' + vv' - \mu v^2v' +  \delta^2 v''' + 2\delta^2v'v'' + \delta^2vv''' -\\
- 4\mu\delta^2 vv'v'' - \mu\delta^2 v'^3 - \mu\delta^2v^2v''' + \frac 25\delta^4v^{(V)} = 0.
\end{gathered}
\end{equation}
Integrating equation \eqref{1d_trav_wave} with respect to $z$ we find that the first integral of equation \eqref{1d_trav_wave} has the form:
\begin{equation}
\begin{gathered}
\label{1d_1stint}
C_1 - C_0v + \frac{1}{2}v^2 - \mu \frac{1}{3} v^3 + \delta^2 v'' + \frac 12\delta^2v'^2 + \delta^2vv'' -\\ -\mu\delta^2vv'^2 - \mu\delta^2v^2v'' + \frac 25 \delta^4v^{(IV)} = 0.
\end{gathered}
\end{equation}
We multiply both sides of equation \eqref{1d_1stint} by $v'$ and integrate the result one more time with respect to z. Then we derive the second integral for equation \eqref{1d_trav_wave}:
\begin{equation}
\begin{gathered}
\label{1d_2ndint}
C_2 + C_1v  - \frac{C_0}2 v^2 +\frac{1}{6}v^3 - \mu \frac{1}{12}v^4 + \frac{\delta^2}2v'^2 + \frac{\delta^2}{2}vv'^2 -\\
- \frac 12 \mu\delta^2v^2v'^2 + \frac 25\delta^4v'v''' - \frac 15 \delta^4 v''^2 = 0.
\end{gathered}
\end{equation}
Hence we have reduced the fifth--order differential equation to the third-order differential equation. It simplifies the process of further investigation.

We apply the Painlev\'e test for equation \eqref{1d_2ndint} using three steps (see for example \cite{Kudryashov2015}). On the first step we have to determine the order of the pole and the first term in expansion of the solution in the Laurent series. We use the equation with the leading terms of equation \eqref{1d_2ndint}:
\begin{equation}
\label{1d_leading_terms}
- \frac 12 \mu\delta^2v^2v'^2 + \frac 25\delta^4v'v''' - \frac 15 \delta^4 v''^2 = 0.
\end{equation}
Substituting $v=\frac{a_0}{z^p}$ into equation \eqref{1d_leading_terms} we find that the general solution of equation \eqref{1d_2ndint} has the pole of the first order ($p=1$) and two branches:
\begin{equation}
v\simeq \pm \frac{4\sqrt{5}\delta}{5\sqrt{\mu}} \frac 1{z}.
\end{equation}
On the second step we determine Fuchs indices. When an equation passes the Painlev\'e test, Fuchs indices indicate the numbers of arbitrary constants in the Laurent expansion of the general solution. To find the Fuchs indices, we substitute expression:
\begin{equation}
v = \pm \frac{4\sqrt{5}\delta}{5\sqrt{\mu}} \frac 1{z} + v_jz^{j-p},
\end{equation}
into equation \eqref{1d_leading_terms}. Equating the linear term with respect to $v_j$ in the expression derived  to zero, we get Fuchs indices for equation \eqref{1d_2ndint}:
\begin{equation}
j_1=-1,\qquad
j_{2,3}=\frac{5\pm i\sqrt{7}}{2}.
\end{equation}
Fuchs indices are the same on both branches of the solution. Two of three indices are complex numbers. It means that equation \eqref{1d_2ndint} does not pass the Painlev\'e test. Thus it  does not have a general solution. However, it is possible to find some exact solutions of equation \eqref{1d_2ndint}. This fact is demonstrated below.

The third step of the test includes examination, whether constants in the Laurent expansion of the solution can be taken as an arbitrary according to the numbers of the Fuchs indices. In our case, this step is not necessary. We also do not need to analyze equation \eqref{1d_final} using the Painlev\'e approach because it is known that if a travelling wave reduction of an equation does not pass the Painlev\'e test, then the equation does not pass it in general case.

We see the analogy to the equations describing the $\alpha$ FPU model. If the high order terms in the Taylor series expansions are not taken into account, then it is possible to describe the model with an integrable equation. This equation admits recurrence of an initial statement and excludes any chaotic behavior. However, if higher order terms are taken into account then it is impossible to derive an integrable equation.

\section{The logistic function method for finding the exact solution of equation \eqref{1d_2ndint}.}
There are various methods to find a solution of a partial differential equation (see for example \cite{polyanin, polyanin2, biswas2, vitanov, vitanov2}). We use the logistic function method. This method often allows to use the results of the Painlev\'e analysis for a partial differential equation for the construction of its exact solutions \cite{Log_func_1, Log_func_2, Log_func_4}. Let us search for the solution of equation \eqref{1d_2ndint} in the following form:
\begin{equation}
\label{1d_log_func_gen}
v = A_0 +  A_1\theta, \quad
\theta = \frac{1}{1-\exp(-k(z-z_0) )},
\end{equation}
where $\theta(z)$ is a so-called logistic function. During the Painlev\'e test, we found the pole order of the general solution. This result is taken into account in expression \eqref{1d_log_func_gen}, where the pole order corresponds to the order found above. We expand  \eqref{1d_log_func_gen} in the Laurent series, substitute it into equation \eqref{1d_2ndint},  equate coefficients at the same powers of $z$ to zero, and solve the derived system of algebraic equations. As a result, we find constants:
\begin{equation}
\begin{gathered}
C_0 = \frac{15-56\mu}{180\mu}, \qquad
C_1 = -\frac{15+56\mu}{360\mu^2},\\
C_2 = -\frac{3136\mu^2+1680\mu+225}{43200\mu^3},
\end{gathered}
\end{equation}
and exact solution of equation \eqref{1d_final}:
\begin{equation}
\begin{gathered}
\label{1d_log_func_solutionA}
v(z) = \pm \sqrt{\frac{2\left(28\mu+15\right)}{15\mu^2}}\left( \frac 1{1-\exp(-k(z-z_0))} -\frac 12 \right) + \frac 1{2\mu}, \\
k = \frac{\sqrt{6\mu\left(28\mu + 15\right)}}{12\mu\delta}.
\end{gathered}
\end{equation}
\begin{figure}
	\centering
	\includegraphics[width=0.7\linewidth]{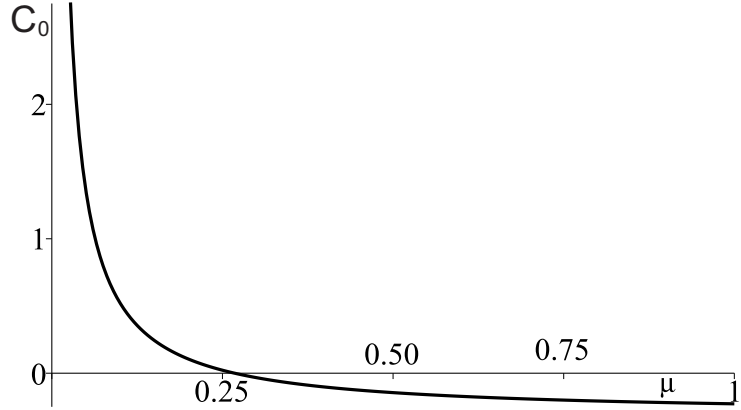}
	\caption{Dependence of the wave velocity $C_0$ from $\mu$}
	\label{fig:velocity}
\end{figure}
The velocity of the wave propagation $C_0$ can be both positive and negative It depends on the correlation of coefficients in the potential of interaction between neighbouring particles. Thus the wave can propagate either to the right or to the left. The velocity as the function of the parameter $\mu$ is presented on figure \ref{fig:velocity}.

Choosing an arbitrary constant $z_0 = z_0 + i\pi/k$ in solution \eqref{1d_log_func_solutionA}, we obtain the solution without poles on the real line:
\begin{equation}
\begin{gathered}
v(z) = \pm \sqrt{\frac{2\left(28\mu+15\right)}{15\mu^2}}\left( \frac 1{1+\exp(-k(z-z_0))} -\frac 12 \right) + \frac 1{2\mu}, \\
k = \frac{\sqrt{6\mu\left(28\mu + 15\right)}}{12\mu\delta}.
\end{gathered}
\end{equation}
Using the following relation:
\begin{equation}
\frac{1}{1+\exp(-k(z-z_0))} = \frac 12 \left(1+\tanh\left(\frac{k(z-z_0)}{2}\right)\right),
\end{equation}
we get the solution of equation \eqref{1d_2ndint} in the form:
\begin{equation}
\begin{gathered}
\label{1d_log_func_solution}
v(z) = \pm \sqrt{\frac{\left(28\mu+15\right)}{30\mu^2}} \tanh\left(\frac{k(z-z_0)}{2}\right) + \frac 1{2\mu}, \\
k = \frac{\sqrt{6\mu\left(28\mu+15\right)}}{12\mu\delta}.
\end{gathered}
\end{equation}
\begin{figure}
	\centering
	\includegraphics[width=0.7\linewidth]{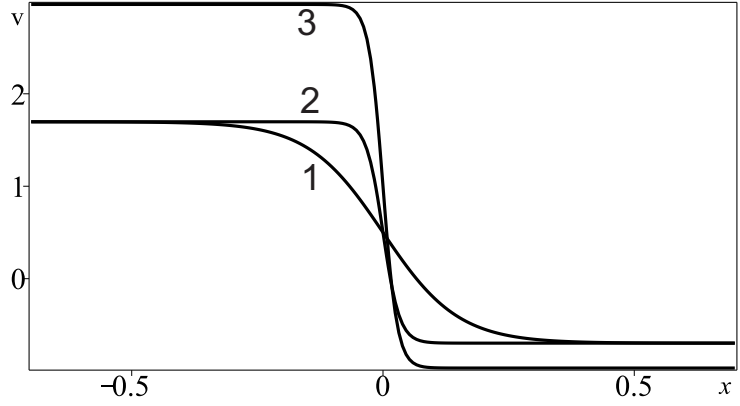}
	\caption{Solutions of equation \eqref{1d_2ndint} at $t=0$ in case of  $\delta=0.1,\; \mu=1$ (line 1), $\delta=0.022,\; \mu=1$ (line 2), and $\delta=0.022,\; \mu=0.5$ (line 3).}
	\label{fig:Log_func_1d_sol}
\end{figure}
Solution \eqref{1d_log_func_solution} has the form of the kink and is illustrated on figure \ref{fig:Log_func_1d_sol} at $t=0$ for different values of parameters. It can be seen from the picture, that the parameter $\delta$ corresponds to the slope of the wave front, while the parameter  $\mu$ is inverse to the amplitude of the perturbation and shifts the solution alongside the ordinate axis.


\section{Elliptic solution for equation \eqref{1d_2ndint}}
Let us use the Laurent expansion method \cite{KudrPLA15, Volkov} to obtain the elliptic solution for equation \eqref{1d_2ndint}. The Laurent expansion method uses the results of the Painlev\'e test as well as the logistic function method. To correlate the pole's order with the order found above, we look for the solution of equation \eqref{1d_2ndint} in the following form \cite{elliptic_Kudr_Dem}:
\begin{equation}
\label{1d_elliptic_solution_gen}
v(z) = H + \frac{A+B\wp_z(z-z_0, g_2, g_3)}{C + \wp(z-z_0, g_2, g_3)},
\end{equation}
where $H,\; A,\; B,\; C$ are some constants, $\wp$ is the elliptic Weierstrass $\wp$-function with invariants $g_2$ and $g_3$. We expand expression \eqref{1d_elliptic_solution_gen} in the Laurent series, substitute this expansion for $v$ in equation \eqref{1d_2ndint}, equate coefficients at the same powers of $z$ to zero, and solve the derived system of algebraic equations. As a result we find the following relations:
\begin{equation}
\begin{gathered}
\label{1d_elliptic_sol_coeff}
H =  \frac{1}{2\mu}á\quad	A = 0,\quad
B = \frac{2\delta}{\sqrt{5\mu}},\quad 
C = -\frac{28\mu+15}{288\mu\delta^2},\\
C_0 = -\frac{2985984\delta^6g_3\mu^3+78400\mu^3+50400\mu^2+10800\mu+3375}{6480\mu^2(15+28\mu)},\\
C_1 = -\frac{2985984\delta^6g_3\mu^3+78400\mu^3+80640\mu^2+27000\mu+3375}{12960\mu^3(15+28\mu)},\\
C_2 = -\frac{336\delta^6g_3}{25\mu} + \frac{1552\delta^6g_3}{25(15+28\mu)} - \frac{343}{150(15+28\mu)}
- \frac {833}{180\mu(15+28\mu)}  -\\
-\frac 7{2\mu^2(15+28\mu)}	-\frac{75}{64\mu^3(15+28\mu)}
+\frac{1296\delta^6g_3}{5\mu(15+28\mu)} -\\
- \frac{75}{512\mu^4(15+28\mu)}  + \frac{3577}{48600\mu} 
+ \frac{721}{4320\mu^2} + \frac{103}{1152\mu^3} + \frac{65}{4608\mu^4}, \\
g_2=-\frac{5971968\delta^6g_3\mu^3-21952\mu^3-35280\mu^2-18900\mu-3375}{20736\delta^4\mu^2(15+28\mu)}.
\end{gathered}
\end{equation}
\begin{figure}
	\centering
	\includegraphics[width=0.7\linewidth]{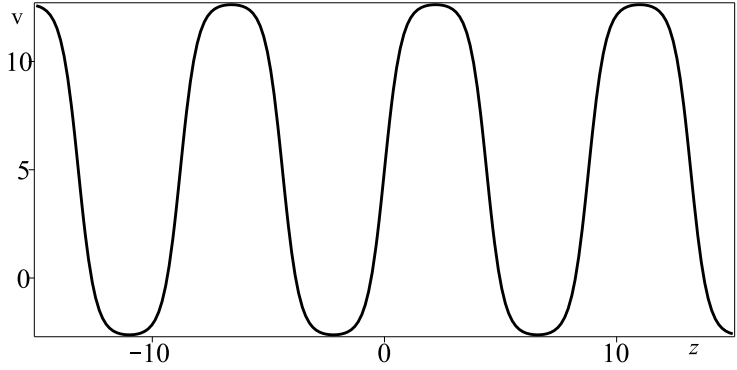}
	\caption{Elliptic solution \eqref{1d_elliptic_solution_gen} for equation \eqref{1d_2ndint} in periodic form in case of $\delta=1,\; \mu=0.5,\; C_0=5$.}
	\label{fig:elliptic}
\end{figure}
Note that $\delta, \mu$, and $g_3$ are parameters in the solution.
One can consider solution \eqref{1d_elliptic_solution_gen} with constants \eqref{1d_elliptic_sol_coeff} and exclude poles from the real line. This solution is presented on figure \ref{fig:elliptic} at $\delta=1,\; \mu=0.5,\; C_0=5$. 

It is useful to investigate the case of the degeneration of the elliptic solution. We take an equation for the $\wp$ function:
\begin{equation}
\label{weierstrass_equation}
y'^2 = 4y^3-g_2y-g_3.
\end{equation}
This equation has the solution $y=\wp (z-z_0, g_2, g_3)$. We explore the right side of equation \eqref{weierstrass_equation} as an algebraic equation. As $g_2=3g_3^{2/3}$ it has roots $y_1=g_3^{1/3},\; y_{2,3} = \frac12 g_3^{1/3}$. In this case, equation \eqref{weierstrass_equation} takes the form:
\begin{equation}
\label{elliptic_eq}
y' = \pm\sqrt{4 \left(y-g_3^{1/3} \right)\left(y+\frac12 g_3^{1/3}\right)^2 }.
\end{equation}
We split the variables and obtain the solution for equation \eqref{elliptic_eq}:
\begin{equation}
\label{y_int}
z-z_0 = \pm \int \frac{dy}{\sqrt{4 \left(y-g_3^{1/3} \right)\left(y+\frac12 g_3^{1/3}\right)^2 } }.
\end{equation}
Considering $y$ as the function of $z$ we derive the solution of equation \eqref{weierstrass_equation} in quadratures:
\begin{equation}
y = g_3^{1/3} + \frac 32 g_3^{1/3} \tan^2 \left( \sqrt {\frac32} g_3^{1/6} \left(z-z_0\right)\right).
\end{equation}
Hence we have the connection between the Weierstrass function in the case of degradation and trigonometric functions. 
\begin{equation}
\wp\left( z-z_0, 3g_3^{2/3}, g_3\right)  =  g_3^{1/3} - \frac 12 g_3^{1/3} \tan^2 \left( \frac1{\sqrt {2}} g_3^{1/3} \left(z-z_0\right)\right).
\end{equation}
It means, that in case of $g_2=3g_3^{2/3}$, an elliptic solution of the differential equation can be expressed in terms of the trigonometric  functions. The period of the elliptic solution converges to infinity. If the poles can be excluded from the real line, then this approach allows to obtain solitary wave solutions from the periodic ones (see for example \cite{Kudr_book}). However for equation \eqref{1d_final} the poles stay on the real line. Thus, solution \eqref{1d_elliptic_solution_gen} with coefficients \eqref{1d_elliptic_sol_coeff} can not be reduced to the form of the solitary wave solution.


\section{Numerical simulation of wave processes, described by equation \eqref{1d_final} with periodic boundary conditions}
We use the pseudospectral method \cite{Kassam2005, Klein2007} for the numerical simulation of the boundary value problem for equation \eqref{1d_final} with periodic boundary conditions. This method was successfully used for the  numerical simulation of different problems of mathematical physics \cite{Log_func_4}. We chose the pseudospectral method because it allows us to solve boundary problems for the equations of high order. This approach is not as demanding to the time step as methods, based on finite difference schemes.

The main idea of the algorithm is the following. We distinguish the linear($L$) and the nonlinear ($N$) operators for  equation \eqref{1d_final}:
\begin{equation}
u(x,t) = L[u]+N[u],
\end{equation}
and introduce the spatial-temporal grid:
\begin{equation}
H =\{0,\; h,\; \cdots, \; N_x\,h; \qquad 0,\; \tau,\; \cdots,\; (n-1)T\},
\end{equation}
where $h$ and $\tau$ are spatial and time steps correspondingly, and $N_x,\; n$ are integer numbers. In each grid point we apply the Fourier space transform:
\begin{equation}
{\hat u_t} = L[\hat{u}] +  N[\hat{u}],
\label{Fourier}
\end{equation}
where $\hat{u} $ is the Fourier transform of the function $u$. The derived differential equation is solved with one of the appropriate methods. We use the fourth-order Runge--Kutta method. 

The smoothing error appears while using the discrete Fourier transform. To correct this the filter is used. The Fourier transform is calculated according to the rule: $\widehat{ \frac{\partial ^n f}{\partial{x^n}} } = (ik_x)^np\hat f$, where $f$ is the transformed function and $p$ is the filter. There are various filters, we use the one that is constructed by the rule of $\frac 23$:
\begin{equation}
p\left( \frac {k_x}{N_x} \right) = \left\{
\begin{gathered}
1, \; \left|{\frac{k_x}{N_x} }\right| > \frac 23 \\
0, \; \left|{\frac {k_x}{N_x} }\right| \le \frac 23.
\end{gathered} \right.
\end{equation}
After that, the original function $u$ is calculated with the inverse Fourier transform.

The Fast Fourier transform is used to accelerate the calculation. It requires spatial grid with the number of grid points equal to the power of 2. Another method to reduce the computational time is the integrating factor method. According to this method, we substitute the unknown function in \eqref{Fourier} in the following way:
\begin{equation}
y = \hat u e^{-\tau\, L}.
\end{equation}
Final formulae for the solution of ODE with the Runge-Kutta method are:
\begin{equation}
\begin{gathered}
a = \tau N\left[\hat  u \right], \\
b = \tau N\left[\left( \hat u + \frac a2 \right)e^{L\tau/2} \right],\\
c = \tau N\left[\hat u e^{L\tau/2} + b/2  \right], \\
d = \tau N\left[\hat u e^{L\tau} + ce^{L\tau/2}  \right], \\
\widehat{u(t+\tau)} = \hat u e^{L\tau} + \frac16 \left( ae^{L\tau} + \left(b+c\right)e^{L\tau/2} + d\right).
\end{gathered}
\end{equation}

\begin{figure}
	\centering
	\includegraphics[width=1\linewidth]{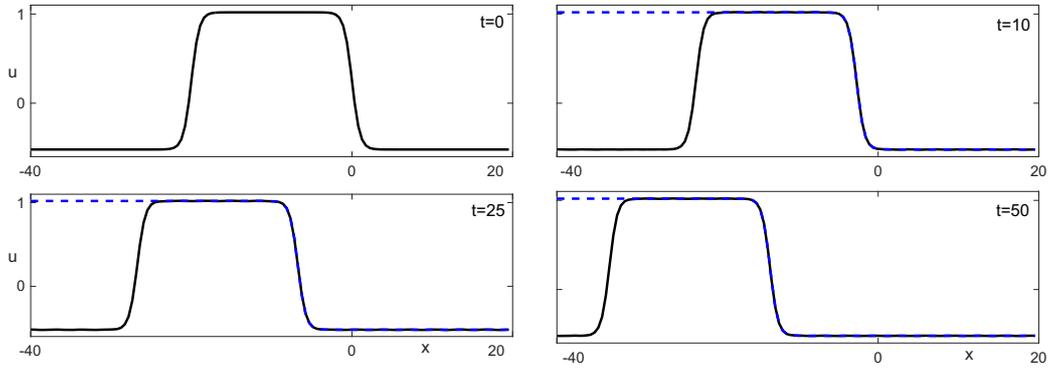}
	\caption{Test of the computational algorithm for exact solution \eqref{1d_log_func_solution} in case of  $\delta=0.6$ and $\mu=2$. The dotted lines are representing the exact solution and the solid lines are for the numerical solution.}
	\label{fig:tanh}
\end{figure}
\begin{figure}
	\centering
	\includegraphics[width=0.7\linewidth]{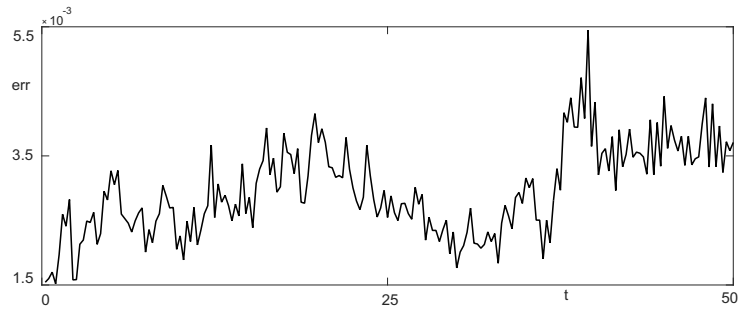}
	\caption{Evolution of the numerical solution error for the boundary value problem for  equation \eqref{1d_final} in case of $\delta=0.6$ and $\mu=2$.}
	\label{fig:error}
\end{figure}
To check our numerical strategy, we use exact solution \eqref{1d_log_func_solution} as initial conditions for the simulation. To keep periodic boundary conditions, we add the reflective part to the exact solution. Hence only a part of the initial conditions is an exact solution for equation \eqref{1d_final}. The result is presented on figure \ref{fig:tanh} at $\delta=0.6$ and $\mu=2$. The solid line is for the numerical solution and the dotted line is for the exact one. It can be seen from the picture, that the solution propagates to the right and does not change its shape. Part of the numerical solution that is obtained from the exact solution coincides with the exact solution. Part of the numerical solution that does not appear to be an exact solution travels without any changes in its shape and velocity or an emission. The computational error was calculated as a ratio $err = \frac{\max|u_{analytic} - u_{calc}|}{\max|u_{calc}|}$, where $u_{analytic}$ and $u_{calc}$ are the exact and the numerical solutions of equation \eqref{1d_final} correspondingly. To calculate $err$ we use only grid points, where the initial conditions for the numerical solution coincides with the exact solution. Evolution of the computational error with time is presented on figure \ref{fig:error}. We see from the figure that the error is below $6\times10^{-3}$, the accuracy of the simulation is quite acceptable.

At $\mu=0$, equation \eqref{1d_final} describes waves in the $\alpha$ FPU model. Solitary wave solutions that look like solitons are found for this case in \cite{KudrPLA15}. We know \cite{boyd} that soliton-like waves described by non-integrable equations usually propagate with radiation. At $\mu \ne 0$, equation \eqref{1d_final} does not have such exact solutions. We can investigate numerically the influence of the additional terms at $\mu \ne 0$  on the stability of such solitary wave solution. We use the solution of equation \eqref{1d_final} at $\mu=0$ from \cite{KudrPLA15}:
\begin{equation}
\begin{gathered}
\label{Sol_KdV5}
w_1(z) = \delta^2\,k^2 - \frac 12 - \frac{3\,\delta^2\,k^2}{2} \tanh^2{\left(\frac{kz}2\right)},\\
z = x + \left(\frac{\delta^4k^4}{10}+\frac 12\right)\,t,
\end{gathered}
\end{equation}
\begin{figure}
	\centering
	\includegraphics[width=0.7\linewidth]{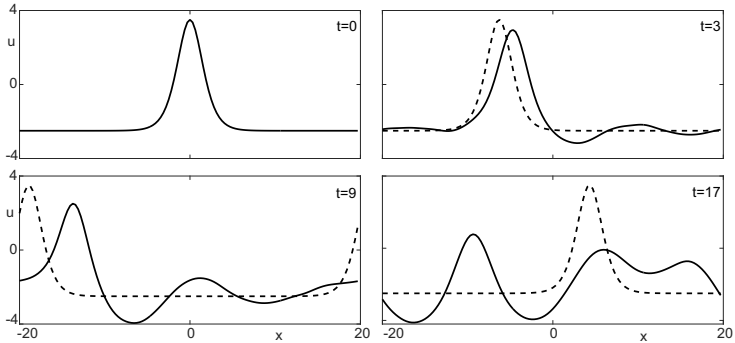}
	\caption{The numerical solution of boundary problem for equation  \eqref{1d_final} in case of $\mu=0$ (dot line) and equation \eqref{1d_final}  in case of $\mu=0.05$ (solid line)  with $k=1,\; \delta=2$.}
	\label{fig:comparis_KDV5}
\end{figure}
as initial conditions. It is reasonably to perform the simulation with small values of the parameter $\mu$. In this case the influence of the quadratic members in the potential of interaction on wave propagation is significantly bigger than the influence of the cubic terms. Figure \ref{fig:comparis_KDV5} illustrates the comparison between numerical solutions of boundary value problems for equation \eqref{1d_final} at $\mu=0$ (the dotted line) and $\mu =0.05$ (the solid line) with $k=1,\; \delta=2$. It can be seen from the reference time that solitary wave slows down and the emission appears at $\mu \ne 0$. Waves created by the emission grow at $t=9$, and the hole destruction of the solitary wave is being observed at $t=17$. At the same time, the numerical solution for boundary value problem at $\mu=0$ changes neither its speed nor its shape. 

\begin{figure}
	\centering
	\includegraphics[width=0.7\linewidth]{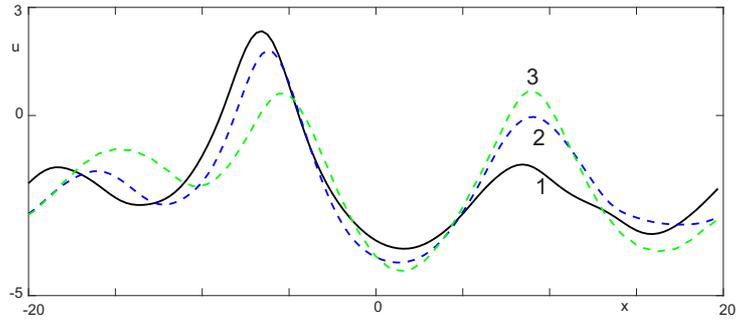}
	\caption{The numerical solution of boundary value problem for equation \eqref{1d_final} at $t=5$ (line 1), $t=27.2$ (line 2) and $t=49.5$ (line 3) in case of $k=1,\; \delta=2$. }
	\label{fig:t5t27,25t49,5}
\end{figure}
\begin{figure}
	\centering
	\includegraphics[width=0.7\linewidth]{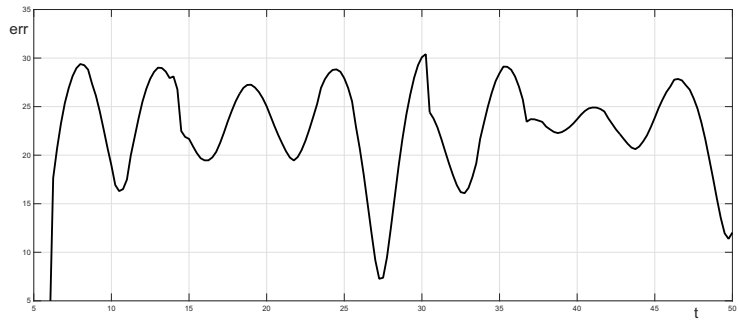}
	\caption{Norm of the difference between the numerical solution for equation \eqref{1d_final} fixed at $t=5$ and current solution.}
	\label{fig:difft5}
\end{figure}

During the simulation one can notice the similar wave profiles at different time moments. The example is presented on figure \ref{fig:t5t27,25t49,5}. We can assume that the recurrence of the perturbations appears as it was observed by Zabusky and Kruskal in work \cite{Kruskal}. To study this, we use the following approach. 
\begin{enumerate}
	\item We fix the numerical solution for equation \eqref{1d_final} in an arbitrary time moment.
	\item We skip a short period of time. Then on each time layer we use the circular convolution method to compare the fixed and current solutions.
	\item We plot the result of calculations as evolution of the absolute value of difference. The difference is calculated in the same way as the computational error.
\end{enumerate}
The result of using this approach for $t=5$ is presented on figure \ref{fig:difft5}. From this figure we can see that minimum of norm of the difference between the two numerical solutions has an obvious period. We can use the program to find this period.
The result is presented in table \ref{tab:period} with an accuracy up to $0.25$. The time moment when the numerical solution was fixed is in the first column of the table. Time moment for the minimum of norm of the difference is in the second column. The third column presents the difference between the first and the second columns. Absolute value of the difference between the solutions at these time moments is in the fourth column. We see from the table that the period of the recurrence is approximately equal to 22.25. Thus that there is no absolute recurrence of the perturbations but we can observe quasi recurrent behaviour. One of the possible reasons for this quasi recurrent dynamics is the short simulation time.
\begin{table}
	\begin{center}
		\begin{tabular}{|c|c|c|c|}
			\hline \rule[-2ex]{0pt}{0.5ex}  t1& t2 & T & diff \\ 
			\hline \rule[-2ex]{0pt}{0.5ex}  5 & 27.25 & 12.25 &  7.2705\\ 
			\hline \rule[-2ex]{0pt}{0.5ex}  10 & 32.25 & 12.25 &  8.6293\\ 
			\hline \rule[-2ex]{0pt}{0.5ex}  15 & 37.25 & 12.25 &  6.3845\\ 
			\hline \rule[-2ex]{0pt}{0.5ex}  20 & 42 & 12 &  6.4039\\ 
			\hline \rule[-2ex]{0pt}{0.5ex}  25 & 47.25 & 12.25 &  6.1192\\ 
			\hline \rule[-2ex]{0pt}{0.5ex}  27.25 & 49.5 & 12.25 &  7.1418\\ 
			\hline \rule[-2ex]{0pt}{0.5ex}  17.5 & 39.75 & 12.25 &  6.1301\\ 
			\hline \rule[-2ex]{0pt}{0.5ex}  19.5 & 41.5 & 12 &  6.6140\\ 
			\hline \rule[-2ex]{0pt}{0.5ex}  21 & 43 & 12 &  6.5962\\ 
			\hline 
		\end{tabular}
		\caption{\label{tab:period} Periods of maximal similarity for the numerical solutions at different time moments.} 
	\end{center}
\end{table}

As we have shown above, equation \eqref{1d_final} extends Gardner equation \eqref{Gardner_eq}. The Gardner equation has solutions in the form of the soliton:
\begin{equation}
\label{Gardner_solution}
u(x,t) = \frac{6C_0}{\cosh\left( \sqrt{\frac{C_0}{\delta^2}}\left(x-C_0t\right)\right) \sqrt{1-6C_0\mu} + 1}.
\end{equation}
\begin{figure}
	\centering
	\includegraphics[width=0.7\linewidth]{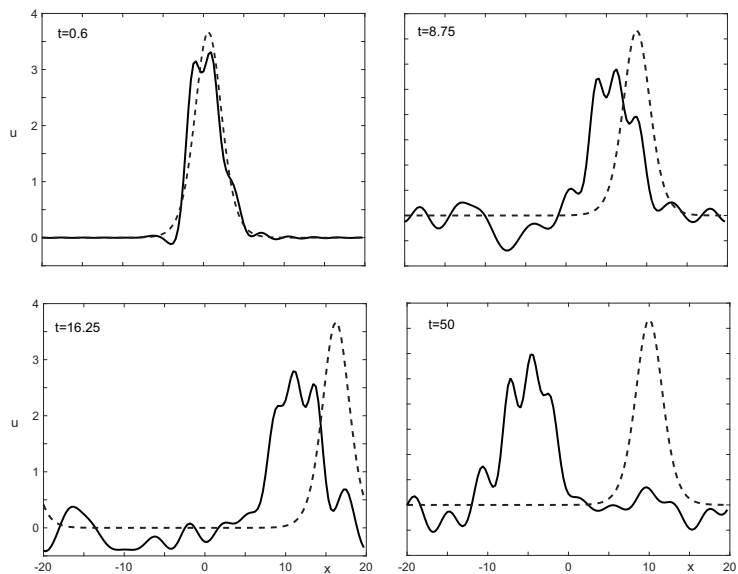}
	\caption{The numerical solutions of boundary problems for equation \eqref{Gardner_eq} (the dotted line) and equation \eqref{1d_final} (the solid line) in case of $\delta=1,\; \mu=0.1,\; C_0=1$. }
	\label{fig:Gardner}
\end{figure}
Let us 	investigate the wave processes, described by equation \eqref{1d_final} with the solitary wave \eqref{Gardner_solution} as initial conditions. The result is presented on figure \ref{fig:Gardner} at $\delta=1,\; \mu=0.1,\; C_0=1$. From the picture we can see that at $t=8.75$ the wave profile is being deformed, creating the burst of waves. This process is accompanied with additional perturbations. The waves are slowing down. Situation is developing at $t=16.25$ and at $t=50$ we can see almost the same picture.

\begin{figure}
	\centering
	\includegraphics[width=0.7\linewidth]{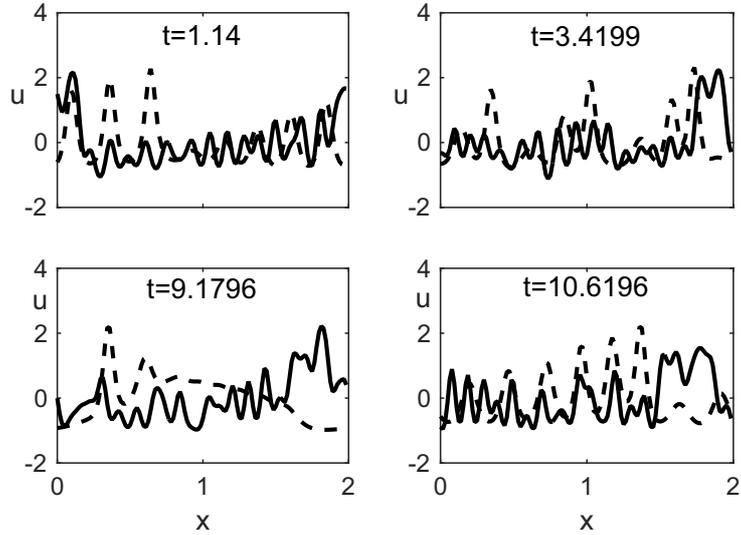}
	\caption{The numerical solutions of boundary problems for the KdV equation (the dotted line) and equation \eqref{1d_final} (the solid line) in case of $\delta=0.022$ and $\mu=1$ }
	\label{fig:cos}
\end{figure}

In paper \cite{Kruskal} the authors provided numerical modelling of wave processes in the FPU mass chain, described by the  Korteweg-de Vries equation. They have conducted the modelling at $\delta=0.022$ in the spatial area  $0 \le x \le 2$. The following initial conditions were used: $u(x,0) = \cos(\pi x)$. We have compared wave processes, described by equation \eqref{1d_final} at $\mu=1$ with the results for the KdV equation using the same initial conditions as in \cite{Kruskal}. The result is illustrated on figure \ref{fig:cos}. The dotted line is for the KdV equation and the solid line is for equation \eqref{1d_final}. We can see from the figure that the numerical solutions of the boundary value problem for the KdV equation consists of solitons. The solitons interact without emission. The solution at $t=10.6$ has the same shape as at $t=1.14$. However, the numerical solution of the boundary value problem for equation \eqref{1d_final} demonstrates a chaotic behaviour. Initial perturbations do not divide into the separated solitons with the elastic interaction. As a result the emission appears and energy leaves initial modes. The recurrence of initial statement can not be observed for wave processes described by equation \eqref{1d_final}.

\section{Conclusion}

We have derived the fifth-order partial differential equation for the description of wave processes in the $\alpha  + \beta$ Fermi--Pasta--Ulam model, taking into account high order terms in the Taylor series expansions. We have shown that the equation does not pass the Painlev\'e test but it has some exact solutions. We have used the logistic function method to find a kink-type solution for the travelling wave reduction of the derived equation. The elliptic solution for the derived equation is constructed with the Laurent expansion method. We have numerically investigated wave processes described by the equation. We have shown that the recurrence of the initial conditions is not specific for an arbitrary initial state. Although we have found a quasi recurrent behaviour of the numerical solution.

\section{Acknowledgments}
This research was supported by Russian Science Foundation Grant no. 14-11-00258.

\end{document}